# Upper critical magnetic field of $LnO_{0.5}F_{0.5}BiS_2$ ($Ln$ = La, Nd) superconductors at ambient and high pressure

Y. Fang[1,2], C. T. Wolowiec[2,3], A. J. Breindel[2,3], D. Yazici[2,3], P.-C. Ho[4], and M. B. Maple[1,2,3*]

[1]*Materials Science and Engineering Program, University of California, San Diego, La Jolla, California 92093, USA*
[2]*Center for Advanced Nanoscience, University of California, San Diego, La Jolla, California 92093, USA*
[3]*Department of Physics, University of California, San Diego, La Jolla, California 92093, USA and*
[4]*Department of Physics, California State University, Fresno, Fresno, California 93740, USA*

The upper critical field $H_{c2}$ of polycrystalline samples of $LnO_{0.5}F_{0.5}BiS_2$ ($Ln$ = La, Nd) at ambient pressure (tetragonal structure) and high pressure (HP) (monoclinic structure) have been investigated via electrical resistivity measurements at various magnetic fields up to 8.5 T. The $H_{c2}(T)$ curves for all the samples show an uncharacteristic concave upward curvature at temperatures below $T_c$, which cannot be described by the conventional one-band Werthamer-Helfand-Hohenberg theory. For the $LaO_{0.5}F_{0.5}BiS_2$ sample under HP, as temperature is decreased, the upper critical field $H_{onset}$, estimated from the onset of the superconducting transitions, increases slowly between 4.9 and 5.8 T compared with the slope of $H_{onset}(T)$ below 4.9 T and above 5.8 T. This anomalous behavior reveals a remarkable similarity in superconductivity between $LaO_{0.5}F_{0.5}BiS_2$ samples measured under HP and synthesized under HP, although the crystal structures of the two samples were reported to be different. The experimental results support the idea that local atomic environment, which can be tuned by applying external pressure and can be quenched to ambient pressure via high temperature-pressure annealing, is possibly more essential to the enhancement of $T_c$ for $BiS_2$-based superconductors than the structural phase transition. On the other hand, such anomalous behavior is very subtle in the case of $NdO_{0.5}F_{0.5}BiS_2$ under HP, suggesting that the anisotropy of the upper critical field in the $ab$-plane and the possible lattice deformation induced by external pressure is weak. This explains why the pressure-induced enhancement of $T_c$ for $NdO_{0.5}F_{0.5}BiS_2$ is not as large as that for $LaO_{0.5}F_{0.5}BiS_2$.

## I. INTRODUCTION

Measurements of the upper critical field, $H_{c2}$, can provide insight into the pair-breaking mechanisms present in superconducting materials and also aid in estimating other characteristics of superconductors such as coherence length and anisotropy. Since the discovery of the cuprates, a large number of high-$T_c$ superconductors have been studied, thus fundamentally challenging the validity of the exsiting Bardeen-Cooper-Schreiffer (BCS) theory of conventional superconductivity. The recently discovered $BiS_2$-based superconducting materials, which exhibit a layered crystal structure similar to the high-$T_c$ cuprates and iron-based superconductors, provide another opportunity to investigate and further understand superconductivity.[1–6] However, current studies on the $BiS_2$-based compounds are only at an early stage and some important questions remain regarding crystal structure, superconductivity, and their interrelation.[7–10]

One of the more striking phenomena displayed by many $BiS_2$-based superconductors, such as $LnO_{1-x}F_xBiS_2$ ($Ln$ = La, Ce, Pr, Nd, Yb), $La_{1-x}Sm_xO_{0.5}F_{0.5}BiS_2$, $Eu_3Bi_2S_4F_4$, $EuBiS_2F$, $SrFBiS_2$, $LaO_{0.5}F_{0.5}BiSe_2$, and $Sr_{0.5}La_{0.5}FBiS_2$, is the rather abrupt enhancement of superconductivity from a low-$T_c$ phase to a high-$T_c$ phase with the application of a moderate amount of pressure on the order of a few GPa.[11–18] X-ray diffraction experiments reveal that $LaO_{0.5}F_{0.5}BiS_2$ and $EuBiS_2F$ undergo a structural phase transition from tetragonal ($P$ $4/nmm$) to monoclinic ($P$ $2_1/m$),[15,19] which is believed to be related to the sudden increase in $T_c$. In addition to the remarkable difference in $T_c$, there are other essential differences in normal state properties between the low-$T_c$ (SC1) and high-$T_c$ (SC2) phases. In particular, the normal state electrical resistivity of $LaO_{0.5}F_{0.5}BiS_2$ in the SC2 phase is significantly smaller than that in the SC1 phase and $NdO_{0.5}F_{0.5}BiS_2$ shows metallic-like behavior in the SC2 phase instead of semiconducting-like behavior as in the SC1 phase.[12,13]

In this paper, we report the evolution of superconductivity under external magnetic fields up to 8.5 T for polycrystalline samples of $LnO_{0.5}F_{0.5}BiS_2$ ($Ln$ = La, Nd) in both the SC1 and SC2 phases. For all samples, the temperature dependence of $H_{c2}$ shows a concave upward curvature with decreasing temperature, which cannot be described by the one-band Ginzburg-Landau theory. The effects of external pressure and chemical composition on superconductivity with increasing external magnetic field for the four samples studied are presented. The anomalous behavior of the temperature dependence of $H_{c2}$, which was observed in the high pressure superconducting phase of $LaO_{0.5}F_{0.5}BiS_2$ and $NdO_{0.5}F_{0.5}BiS_2$ at ~5 T and ~3 T, respectively, will be described, and a comparison of the superconductivity observed in samples of $LaO_{0.5}F_{0.5}BiS_2$ in the high pressure phase (SC2) and the superconductivity observed in the high-pressure synthesized samples of $LaO_{0.5}F_{0.5}BiS_2$ which are measured at ambient pressure, will be discussed. We believe the results will be useful in understanding (1) whether the $BiS_2$-based superconductors are conventional or unconventional and (2) how the difference in crystal structure and local atomic environment of the the $BiS_2$-based compounds affects superconductivity. This study also provides the first possible explanation that we are aware of for why pressure-induced enhancements of $T_c$ for $LnO_{0.5}F_{0.5}BiS_2$ ($Ln$ = La-Nd) decrease with heavier $Ln$.

For convenience in the following discussion, as-grown samples measured at ambient pressure and under high pressure will be abbreviated as AG and HPAG, respectively. In addition, samples that were synthesized and studied by Mizuguchi *et al.* (Ref. 20), which were grown under high-pressure and high-temperature conditions, shall be abbreviated as HPT.

## II. EXPERIMENTAL DETAILS

As-grown polycrystalline samples of $LnO_{0.5}F_{0.5}BiS_2$ ($Ln =$ La, Nd) were synthesized and annealed at ~800 °C in sealed quartz tubes as described elsewhere.[4,6] The AG and HPAG samples of the same chemical composition came from the same pellet to ensure both samples have the same physical properties at ambient pressure. Geometric factors used in determining the resistivity for each sample were measured before applying pressure. Hydrostatic pressure was generated by using a clamped piston-cylinder cell (PCC) in which a 1:1 by volume mixture of n-pentane and isoamyl alcohol was used as the pressure-transmitting medium. The pressures applied to the samples were estimated by measuring the $T_c$ of a high purity (>99.99%) Sn disk inside the sample chamber of the cell and comparing the measured values with the well-determined $T_c(P)$ of high purity Sn.[21] The resistivity measurements of the AG LaO$_{0.5}$F$_{0.5}$BiS$_2$ sample at magnetic fields up to 1 T were performed by using a Quantum Design Physical Property Measurement System (PPMS) Dynacool using a standard four-wire technique from 5 to 1.8 K. Electrical resistivity measurements at temperatures down to 60 mK on the HPAG LaO$_{0.5}$F$_{0.5}$BiS$_2$ sample and the AG and HPAG NdO$_{0.5}$F$_{0.5}$BiS$_2$ samples were performed with an Oxford Instruments Kelvinox $^3$He-$^4$He dilution refrigerator in magnetic fields ranging from 0 to 8.5 T. Previous studies have shown that the pressure-induced phase transition for LaO$_{0.5}$F$_{0.5}$BiS$_2$ and NdO$_{0.5}$F$_{0.5}$BiS$_2$ occurs at around 0.8 and 1.9 GPa, respectively, followed by a gradual decrease in $T_c$ in both compounds with additional pressure.[12,13,19] The HPAG samples were measured under an applied pressure of ~2.3 GPa in the clamped PCC, which was attached to the dilution refrigerator probe. The value of 2.3 GPa, which is beyond the phase transition pressure of 1.9 GPa for NdO$_{0.5}$F$_{0.5}$BiS$_2$, was chosen to ensure that the SC2 phase was fully realized in the HPAG samples.

## III. RESULTS AND DISCUSSION

Electrical resistivity $\rho$ vs. temperature $T$ at various magnetic fields up to 8.5 T in the temperature range down to 60 mK for LaO$_{0.5}$F$_{0.5}$BiS$_2$ (AG and HPAG) and NdO$_{0.5}$F$_{0.5}$BiS$_2$ (AG and HPAG) is displayed in Fig. 1(a)-(d), respectively. The normal state electrical resistivity $\rho$ increases slightly with increasing magnetic field in the normal state for all four samples, revealing a positive magnetoresistivity. Superconductivity for all the samples is suppressed by the external magnetic field, as evinced by the shift of the superconducting transition curves to lower temperature with increasing magnetic field. If we define $T_c$ as the temperature at which $\rho$ falls to 50% of its normal state value, the $T_c$ values for LaO$_{0.5}$F$_{0.5}$BiS$_2$ and NdO$_{0.5}$F$_{0.5}$BiS$_2$ at 2.3 GPa are 8.28 K and 6.31 K, respectively, and the $T_c$ values for the corresponding AG samples are 2.92 K and 4.54 K at zero magnetic field. The large difference in $T_c$ between the AG and HPAG samples of the same compound and the sharp superconducting transition at zero magnetic field reveals that the AG samples are in the SC1 phase and the HPAG samples are in the SC2 phase.

To analyze the evolution of superconductivity as a function of magnetic field, we plotted the $T$ dependence of the upper critical field in terms of the characteristic fields $H_{10\%\rho}$ and $H_{90\%\rho}$, evaluated at 10% and 90% of the normal state $\rho$ at the onset of the superconducting transition in Figs. 2(a) and 2(b), respectively. At high magnetic fields, some of the $H_{10\%\rho}(T)$ data are not shown since the temperatures are not low enough to reach the 10% value of its normal state resistivity. As shown in Figs. 1 and 2, superconducting transitions for all the samples are quite broad in high magnetic fields. It has been reported that anisotropy in single-crystalline samples of BiS$_2$-based superconductors is quite large.[22-24] For example, the upper critical field parallel to the $ab$-plane ($H_{c2}{}^{ab}$) for single-crystalline NdO$_{0.5}$F$_{0.5}$BiS$_2$ is estimated to be ~42 T, whereas the upper critical field parallel to the $c$-axis ($H_{c2}{}^c$) for the same compound is only about 1.3 T.[22] In other words, the superconducting state of grains whose $c$-axis is parallel to the applied magnetic field is more rapidly suppressed by applying high magnetic fields than grains whose $ab$-plane is parallel to the applied magnetic field.[20] In this study, the measured magnetization curves are reversible, suggesting weak vortex pinning. For the polycrystalline samples whose grains are randomly distributed and separated by grain boundaries and small pores, the significant difference between $H_{90\%\rho}(T)$ and $H_{10\%\rho}(T)$ is expected to be related to the large anisotropy of the upper critical field; $H_{90\%\rho}$ is then expected to be close to the upper critical field for grains whose $ab$-plane is parallel to $H$.[20] The difference in $T$ between the $H_{10\%\rho}(T)$ and $H_{90\%\rho}(T)$ points on the transition curve for a particular magnetic field can be regarded as the width of superconducting transition at that field.

Application of an external magnetic field may destroy the Cooper pairs via the pair breaking interaction between the magnetic field and/or the momenta of the electrons (electromagnetic interaction) and the spins of the electrons (Zeeman interaction). With increasing $H$, superconductivity is suppressed more rapidly for the AG samples compared to the corresponding HPAG samples with the same chemical composition. Superconductivity in AG LaO$_{0.5}$F$_{0.5}$BiS$_2$ cannot be observed above 1.9 K when $H$ reaches 1 T (shown in Fig. 1(a)). As can be seen in Fig. 2(a), despite the difference in $T_c$ at zero magnetic field, the $H_{10\%\rho}(T)$ curve for HPAG LaO$_{0.5}$F$_{0.5}$BiS$_2$ is almost parallel to the $H_{10\%\rho}(T)$ curve for HPAG NdO$_{0.5}$F$_{0.5}$BiS$_2$ up to 8.5 T, while the $H_{90\%\rho}(T)$ curves for the two samples have noticeably different slopes where $dH/dT$ for HPAG NdO$_{0.5}$F$_{0.5}$BiS$_2$ is steeper than $dH/dT$ for HPAG LaO$_{0.5}$F$_{0.5}$BiS$_2$. Similar behavior can be found in AG samples of LaO$_{0.5}$F$_{0.5}$BiS$_2$ and NdO$_{0.5}$F$_{0.5}$BiS$_2$ that are in the same SC1 phase as shown in the inset of Fig. 2(b) where the $H_{10\%\rho}(T)$ curves track each other closely. However, a comparison of the upper critical field curves shown in Fig. 2(a) and Fig. 2(b), indicates that the temperature dependence of $H_{10\%\rho}$ and $H_{90\%\rho}$ for samples (of the same lanthanide constituent) in the SC1 phase is more gradual than that of samples in the SC2 phase. The differences observed in the upper critical field behavior between the SC1 and SC2 phases suggest the superconducting properties of these materials change significantly during the structural phase transitions.

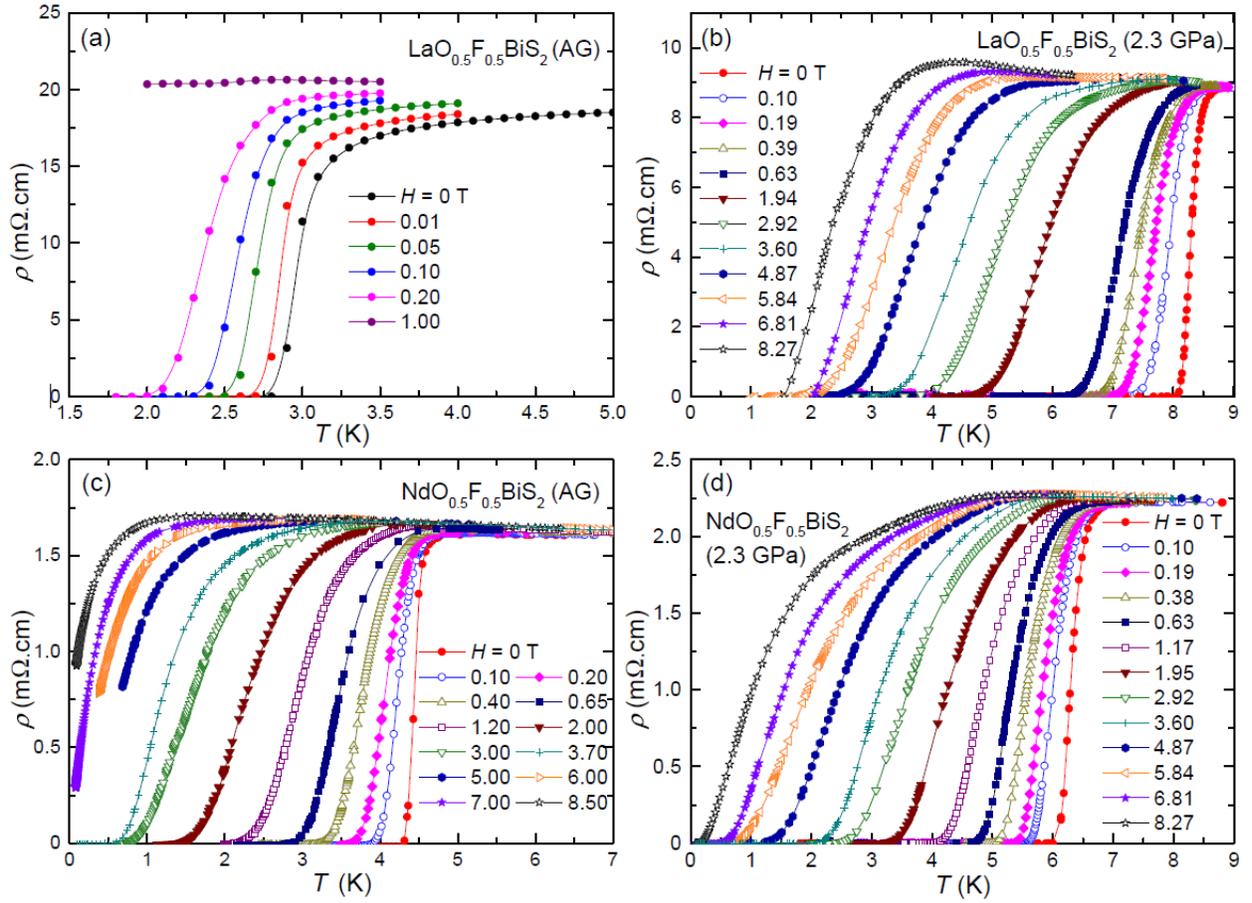

FIG. 1. (Color online) (a), (b) Temperature $T$ dependence of the electrical resistivity $\rho$ for AG and HPAG samples of LaO$_{0.5}$F$_{0.5}$BiS$_2$, respectively. (c), (d) Electrical resistivity $\rho$ vs. temperature for AG and HPAG samples of NdO$_{0.5}$F$_{0.5}$BiS$_2$, respectively. The external magnetic field for each $\rho(T)$ curve is also denoted in the figure.

In the conventional one-band Ginzburg-Landau picture, the upper critical field increases linearly with decreasing $T$ near $T_c$ and then saturates to a finite value in the 0 K limit. However, as can be seen in Fig. 2, the $T$ dependence of both $H_{90\%\rho}$ and $H_{10\%\rho}$ shows an uncharacteristic upward curvature with decreasing $T$ down to the lowest temperature reached in this study. This is different from the upper critical field behavior described by the standard one band model of Werthamer-Helfand-Hohenberg (WHH).[25] Similar behavior was also reported for AG samples of LaO$_{1-x}$F$_x$BiS$_2$ ($x = 0.1$-$0.3$) and Bi$_4$O$_4$S$_3$ based on electrical resistivity measurements.[26,27] However, other studies on

CeO$_{0.5}$F$_{0.5}$BiS$_2$, Bi$_4$O$_4$S$_3$, La(O,F)BiSSe, PrO$_{1-x}$F$_x$BiS$_2$, Sr$_{0.5}$La$_{0.5}$FBiS$_2$ show nearly linear $T$ dependences of $H_{c2}$ that can be described by Ginzburg-Landau theory.[28–32] The lowest temperature reached in these studies was ~2 K, and it should be mentioned that the range of values for the superconducting transition temperatures for the BiS$_2$-based compounds at zero magnetic field is 2-11 K. In the present study, the upper critical field in the vicinity of $T_c$ in zero magnetic field is nearly linear in $T$ and can be fitted with the WHH equation. Since the chemical compositions and crystal structures of the BiS$_2$-based compounds are very similar, it is possible that the previously reported $H_{c2}$ vs. $T$ curves would also deviate from the WHH model when extended to lower temperature.

In the neighborhood of $T_c$, the upper critical field estimated by applying the one-band WHH equation is quite close to the experimental data; however, the difference becomes more significant with decreasing $T$. For example, as temperature goes to zero K, the one band WHH model for $H_{90\%\rho}$ deviates significantly from the experimental data obtained for AG NdO$_{0.5}$F$_{0.5}$BiS$_2$ (shown in Fig. 2(b)). While the model extrapolates to a value of 4.2 T at $T = 0$ K, the data climbs to a field of more than twice that value as the temperature ap-proaches 0 K. The significant low-$T$ enhancement of $H_{c2}(T)$ may indicate multi-band superconductivity for BiS$_2$-based compounds. In the present work, we adopted the dirty two band model introduced in Ref. 33 (Eq. 19), which treats both interband scattering and paramagnetic effects as negligible. The two band model was fitted to the data and plotted in Fig. 2(b) and the fitting $H_{90\%\rho}(T)$ and $H_{10\%\rho}(T)$ curves for the AG samples of NdO$_{0.5}$F$_{0.5}$BiS$_2$ resulted in estimated values of $H_{90\%\rho}(0) = 9.1$ T and $H_{10\%\rho}(0) = 4.5$ T, respectively. It should be noted that the $H_{c2}(0)$ value of 16 T ($H$ parallel to $ab$-plane) estimated by using WHH theory reported on single crystalline NdO$_{0.5}$F$_{0.5}$BiS$_2$ is significantly higher than the $H_{90\%\rho}(0)$ value of 9.1 T reported in this work. The concave upward curvature of $H_{c2}(T)$ and the different fitting models should be responsible for causing the difference of the values.

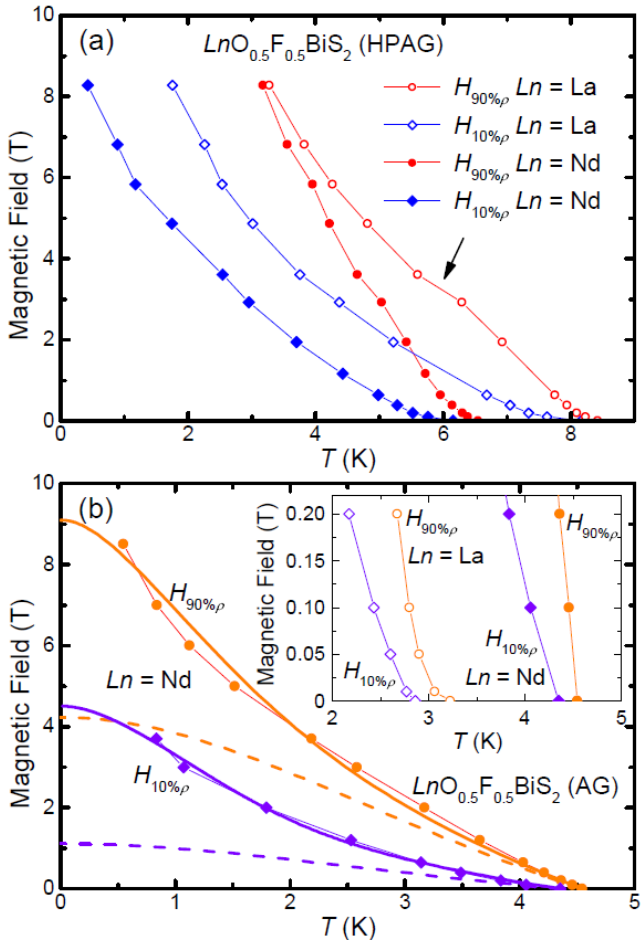

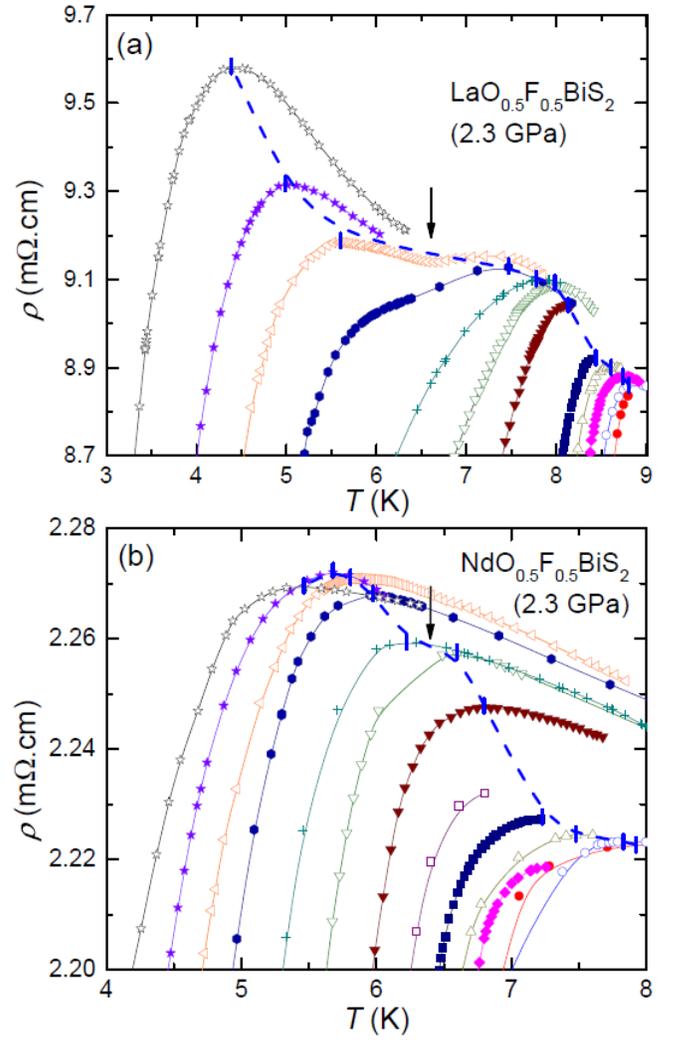

FIG. 2. (Color online) (a) $H_{90\%\rho}$ and $H_{10\%\rho}$ for the samples in the SC2 phase plotted as a function of $T$. The open and filled symbols represent data obtained from LaO$_{0.5}$F$_{0.5}$BiS$_2$ and NdO$_{0.5}$F$_{0.5}$BiS$_2$, respectively. (b) $T$ dependence of $H_{90\%\rho}$ and $H_{10\%\rho}$ for AG NdO$_{0.5}$F$_{0.5}$BiS$_2$. The dashed and solid lines are fits to the experimental data (circles for $H_{90\%\rho}$ and diamonds for $H_{10\%\rho}$) of the one-band WHH equation and the two band model, respectively. The inset shows $H_{90\%\rho}$ and $H_{10\%\rho}$ values (circular and rhombic symbols, respectively) for the AG LaO$_{0.5}$F$_{0.5}$BiS$_2$ (open symbols) and NdO$_{0.5}$F$_{0.5}$BiS$_2$ (filled symbols) from 0 to 0.20 T.

FIG. 3. (Color online) Enlargement of $T$ dependence of $\rho$ for (a) HPAG LaO$_{0.5}$F$_{0.5}$BiS$_2$ and (b) HPAG NdO$_{0.5}$F$_{0.5}$BiS$_2$ near the onset of the superconducting transition. Black arrows indicate the anomalous behavior of $H_{onset}(T)$. The vertical ticks represent the onset of superconducting transition ($T_c^{onset}$) and the dashed lines are guides to the eye.

It should be mentioned that although the two band model provides a better overall fit and a more reasonable $H_{90\%\rho}(0)$ value (~9.1 T for AG NdO$_{0.5}$F$_{0.5}$BiS$_2$) compared with the one-band WHH model, the model is still unable to account for the strong upward curvature exhibited in the experimental data at low temperatures. Other than the possibility of multi-band superconductivity (three or more bands), there are several possible mechanisms that may be responsible for the significantly higher values of $H_{c2}$ at low temperatures, including a crossover from three-dimensional to two-dimensional behavior in layered superconductors, and anisotropy in the superconducting gap and/or Fermi velocity.[26,34,35] However, no conclusive evidence has been reported to support these mechanisms. It should be also noted that previous experimental and theoretical studies cannot reach an agreement of whether the newly discovered BiS$_2$-based superconductors are conventional or unconventional. The poor description of the $H_{c2}(T)$ data is possibly a result of unconventional superconductivity. To investigate the origin of the significant upward curvature of $H_{c2}$ with decreasing $T$, further studies of the superconductivity of BiS$_2$-based compounds are needed.

For the HPAG LaO$_{0.5}$F$_{0.5}$BiS$_2$ sample, the increase of $H_{90\%\rho}$ with decreasing $T$ slows down somewhat when the external magnetic field reaches ~3 T (indicated by the black arrow in Fig. 2(a)) and then recovers to its lower field rate above 4 T. This anomalous behavior is more remarkable in the evolution of $T_c^{onset}$, which is defined as the $T$ at which $\rho$ attains its maximum value, in various external magnetic fields. As shown in Fig. 3(a), $T_c^{onset}$ is significantly suppressed with a slight increase in $H$ at around 5 T. The temperature dependence of $H_{onset}$ for HPAG LaO$_{0.5}$F$_{0.5}$BiS$_2$ is shown in Fig. 4, where the values of $T_c^{onset}$ were determined from $\rho(T)$ and are depicted by the short vertical solid lines as displayed in Fig. 3. From 0.6 to 4.8 T, $H_{onset}(T)$ increases almost linearly with decreasing $T$ with a slope $dH_{onset}/dT$ of -4.2 T/K; however, the magnitude of $|dH_{onset}/dT|$ decreases to 0.52 T/K in the range from 4.8 to 5.9 T and then increases to 2.4 T/K above 6.0 T. The $H_{onset}(T)$ curve for HPAG LaO$_{0.5}$F$_{0.5}$BiS$_2$ is remarkably similar to that observed in the HPT sample of LaO$_{0.5}$F$_{0.5}$BiS$_2$ which was synthesized at a pressure of 2 GPa and temperature of 700 °C (see Fig. 4).[20] Possible anisotropy of the superconducting state within the BiS$_2$ plane, which may be induced by lattice distortion or residual strain along the superconducting layers in the high pressure and high temperature condition, is considered to be the origin of the $H_{onset}(T)$ difference below and above 8 T.[20] In other words, the upper critical field for HPT LaO$_{0.5}$F$_{0.5}$BiS$_2$ synthesized at 2 GPa is different along the $a$- and $b$- axes. In this study, although the samples under high pressure are still considered to be in a good hydrostatic environment, anisotropy of the upper critical field along the $a$- and $b$-axes can be induced due to the pressure-induced phase transition from the tetragonal phase (space group $P\,4/nmm$) to the monoclinic phase (space group $P\,2_1/m$).

Since the anomalous behavior in $H_{onset}(T)$ is very remarkable in HPAG LaO$_{0.5}$F$_{0.5}$BiS$_2$, such anomalous behavior may also be observed in the HPAG NdO$_{0.5}$F$_{0.5}$BiS$_2$ due to the similarity in chemical composition and crystal structure between the two compounds. As displayed in Fig. 3(b) (black arrow) and Fig. 4 (red arrow), the $T$ dependence of $H_{onset}$ for HPAG NdO$_{0.5}$F$_{0.5}$BiS$_2$ shows a very subtle discontinuity in a narrow $T$ range ~0.5 K at around 3 T. It should be mentioned that a variety of theoretical and experimental studies have suggested that the band structure and superconductivity of the BiS$_2$-based compounds can be largely affected by a change in the crystal structures.[5,7,36–40] Anisotropic superconducting states within the $ab$-plane could be induced when the degeneracy of Bi-6$p_x$ and Bi-6$p_y$ orbital is lifted.[20] Based on the previous discussions, the results indicate that the anisotropy of the upper critical field in the $ab$-plane of HPAG NdO$_{0.5}$F$_{0.5}$BiS$_2$ is not as large as that of HPAG LaO$_{0.5}$F$_{0.5}$BiS$_2$, and hence, suggests that the pressure-induced lattice deformation at 2.3 GPa in NdO$_{0.5}$F$_{0.5}$BiS$_2$ is not significant compared with that in LaO$_{0.5}$F$_{0.5}$BiS$_2$. This also explains why the pressure-induced enhancement of the superconducting critical temperature $\Delta T_c$ in NdO$_{0.5}$F$_{0.5}$BiS$_2$ is only ~2.5 K, which is much lower than the $\Delta T_c$ value of LaO$_{0.5}$F$_{0.5}$BiS$_2$ (~7.2 K).

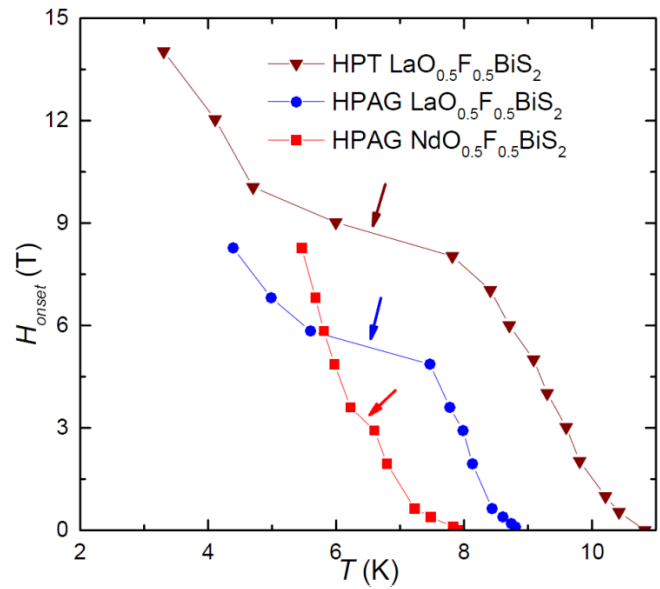

FIG. 4. (Color online) (a) The $H_{onset}$-$T$ phase diagrams for HPAG and HPT LaO$_{0.5}$F$_{0.5}$BiS$_2$ as well as HPAG NdO$_{0.5}$F$_{0.5}$BiS$_2$. The data for HPT LaO$_{0.5}$F$_{0.5}$BiS$_2$ are taken from Ref. 20.

Previous studies have revealed that it is possible to significantly increase $T_c$ in the $Ln$(O,F)BiS$_2$ ($Ln$ = La-Nd) superconductors by applying high pressure or by annealing/synthesizing the samples under high pressure.[12,13,18,19,41] Although the crystal structure of HPT samples is reported to be the same as the structure of AG samples at low pressure and is different from the monoclinic structure of the same compounds under high pressure,[41–43] $T_c$ values of HPT samples and the corresponding (same chemical composition) HPAG samples in SC2 phase are very close. The enhancement of $T_c$ for these compounds under pressure is considered to be related to the structural phase transition; however, there is no conclusive agreement regarding the enhancement of superconductivity in samples that are annealed or synthesized under pressure.[7,42–45] The $T_c$ value of HPAG LaO$_{0.5}$F$_{0.5}$BiS$_2$ plotted in Fig. 3 is lower than the reported $T_c$ value of the HPT LaO$_{0.5}$F$_{0.5}$BiS$_2$. This discrepancy can be attributed to the fact that the measurement of electrical resistivity on the HPAG sample of LaO$_{0.5}$F$_{0.5}$BiS$_2$ was performed at 2.3 GPa, well into the SC2 phase, where a gradual suppression of $T_c$ occurs with increasing pressure. Nevertheless, the values of $dH_{onset}/dT$ are nearly the same at each of the three different stages in the evolution of $H_{onset}(T)$ as shown in Fig. 4. Our results raise the question of why superconductivity for the same compounds in this system, enhanced by the two different methods, is so similar. A thorough investigation of this problem may yield information that will help identify the essential parameters that determine $T_c$ in the BiS$_2$-based superconductors.

One possible explanation for the enhanced superconductivity observed in the HPT samples of $Ln$O$_{0.5}$F$_{0.5}$BiS$_2$ ($Ln$

= La – Nd) is the presence of trace amounts of the monoclinic structure at ambient pressure; however, it has been observed that the pressure dependence of $T_c$ for the HPAG samples is completely reversible, whereby $T_c$ was determined from measurements of electrical resistivity upon a gradual and stepwise unloading of the pressure from the piston-cylinder pressure cell. This suggests it is unlikely for the monoclinic phase to survive in returning from high pressure to atmospheric pressure.[12,13,18] In addition, if these samples are mixtures of two phases with distinct $T_c$ values, there is possibly additional curvature in the $T$ dependence of the resistivity, magnetic susceptibility, or specific heat due to the appearance of superconductivity from an additional superconducting phase. Such a broadening of the resistive superconducting transitions might look similar to the broadened $\rho(T)$ curves observed in the vicinity of the phase transition pressures for $Ln$O$_{0.5}$F$_{0.5}$BiS$_2$ ($Ln$ = La, Ce, Pr, Nd) and La$_{1-x}$Sm$_x$O$_{0.5}$F$_{0.5}$BiS$_2$ under external pressure. It has also been suggested that the enhanced superconductivity in the HPT samples may result from additional effects that high pressure annealing may have on the local crystal structure including the shorter in plane Bi–S distances and higher symmetry in the $ab$-plane reported for the Ce(O,F)BiS$_2$ compound as well as the uniaxial strain along the $c$-axis that was observed in HPT samples of LaO$_{0.5}$F$_{0.5}$BiS$_2$ and PrO$_{0.5}$F$_{0.5}$BiS$_2$.[42–45]

The pressure-induced phase transition observed in LaO$_{0.5}$F$_{0.5}$BiS$_2$ was reported to involve sliding between the two neighboring BiS$_2$ layers along the $a$-axis, resulting in a slight increase of the angle between the $ab$ and $bc$ planes ($\beta$) from 90° at ambient pressure to 94° at ~1 GPa.[19] The in-plane structure of the BiS$_2$ layers, which is considered to be essential for the superconductivity, as well as the in-plane structure of the La(O,F) blocking layers, are nearly the same after the phase transition. Considering the results obtained in this study, it seems that local distortions or changes in the local atomic environment, as caused by the application of pressure but also quenched via high pressure annealing, are probably critical in affecting superconductivity, perhaps even more than the structural phase transition itself. However, further investigations of the crystal structure of the AG, HPAG, and HPT are still needed for direct evidence of changes in local structure. Very recently, it was reported that "in-plane chemical pressure" is closely related to $T_c$ in the BiS$_2$-based superconductors.[39] The in-plane chemical pressure is defined as the ratio of the expected bond distance between a Bi ion and its in-plane neighboring ions of S (or Se) to the experimental bond distance estimated by Rietveld refinements. The expected bond distance can be determined by using the ionic radii of Bi and S (or Se). Although the validity of such a relationship needs to be further confirmed, the results also emphasize the importance of local structure to superconductivity in BiS$_2$-based compounds.

## IV. SUMMARY

To summarize, we performed electrical resistivity measurements on polycrystalline samples of $Ln$O$_{0.5}$F$_{0.5}$BiS$_2$ ($Ln$ = La, Nd) in both the SC1 and SC2 phases under external magnetic fields up to 8.5 T and at temperatures ranging from 60 mK to 11 K. Superconducting transitions became much broader and the values of $T_c$ were suppressed for all the samples under high magnetic field. Significant concave upward curvatures in the $H_{c2}(T)$ curves were observed and cannot be described by conventional one-band WHH theory. In addition, the $T$ dependence of $H_{onset}$ for HPAG LaO$_{0.5}$F$_{0.5}$BiS$_2$ shows anomalous behavior at around 5 T, revealing remarkable similarity to the superconductivity observed the HPT samples of LaO$_{0.5}$F$_{0.5}$BiS$_2$. If no high-pressure monoclinic phase exists in HPT LaO$_{0.5}$F$_{0.5}$BiS$_2$, the importance of the monoclinic phase would be diminished, suggesting the possibility of a greater importance in the role that local atomic environment plays in affecting superconductivity in the BiS$_2$-based compounds, though further investigations of atomic positions are still needed. In the case of HPAG NdO$_{0.5}$F$_{0.5}$BiS$_2$, the anomalous behavior in $H_{onset}(T)$ is very subtle, probably due to a mild pressure-induced deformation in crystal structure, which is probably responsible for the relatively slight increase in $T_c$ compared to the remarkable enhancement of $T_c$ in LaO$_{0.5}$F$_{0.5}$BiS$_2$.

## ACKNOWLEDGMENTS

High pressure research at UCSD was supported by the National Nuclear Security Administration under the Stewardship Science Academic Alliance Program through the US Department of Energy under Grant No. DE-NA0002909. Materials synthesis and characterization at UCSD was supported by the US Department of Energy, Office of Basic Energy Sciences, Division of Materials Sciences and Engineering, under Grant No. DEFG02-04-ER46105. Low temperature measurements at UCSD were sponsored by the National Science Foundation under Grant No. DMR 1206553. Research at CSU-Fresno was supported by the National Science Foundation under Grant No. DMR-1506677.

---